\documentclass[preprint,preprintnumbers,showpacs,aps,amssymb]{revtex4-1}

\usepackage{graphicx}
\usepackage{bm}
\usepackage{amsmath}

\def\calO{{\cal O}}
\def\calU{{\cal U}}

\def\dU{d_{\calU}}
\def\MPl{M_{\rm Pl}}
\def\dUV{d_{\rm UV}}

\begin{document}
\title{Quasinormal modes of unparticle black holes}
\author{Jong-Phil Lee}
\email{jplee@kias.re.kr}
\affiliation{Department of Physics and IPAP, Yonsei University, Seoul 120-749, Korea}
\affiliation{Division of Quantum Phases $\&$ Devices, School of Physics, Konkuk University, Seoul 143-701, Korea}

\begin{abstract}
In the background of unparticle-enhanced black hole geometry, we provide the quasinormal modes of scalar, vector, and
Dirac particles around it.
Ungravity by tensor unparticles contributes positively to the Newtonian gravity and black holes can be formed at the LHC
without any extra dimensions.
In the ungravity-dominant regime, the gravity looks much like that of the Schwarzschild geometry in fractional number of extra dimensions.
We argue in this analysis that quasinormal modes are good fingerprints to distinguish ungravity from extra dimensions,
by showing that the unitarity constraints on ungravity forbid some of the quasinormal modes which are allowed in extra dimensions.
\end{abstract}

\maketitle
\section{Introduction}
With the successful running of the Large Hadron Collider (LHC) at CERN, we are entering a new age of high energy physics.
It is highly anticipated that the last piece of the Standard Model (SM), the Higgs boson, would be discovered or ruled out 
about within a year, and new physics signals beyond the SM would appear soon.
Among the new physics is the unparticle \cite{Georgi}.
But it is quite different from other types of new physics.
Many kinds of new physics --- supersymmetry or extra dimensions, etc. --- involve some new sets of {\em particles}, 
while unparticles are not ordinary particles.
In the unparticle scenario, there is a scale-invariant hidden sector which couples to the SM particles very weakly at
some energy scale $\Lambda_\calU$.
When seen at low energy, the scale-invariant sector behaves in different ways from ordinary particles and it looks like a
fractional number of particles, and hence dubbed as {\em unparticles}.
\par
Suppose that at some high energy $\sim M_\calU$, 
there is a ultraviolet (UV) theory in the hidden sector with the infrared (IR)-stable fixed point.
The interaction between the UV theory and the SM sector can be described by an effective theory formalism.
Below $M_\calU$, a UV operator $\calO_{\rm UV}$ interacts with an SM operator $\calO_{\rm SM}$ through 
$\calO_{\rm SM}\calO_{\rm UV}/M_\calU^{d_{\rm SM}+d_{\rm UV}-4}$.
Here $d_{\rm UV(SM)}$ is the scaling dimension of $\calO_{\rm UV(SM)}$.
Through the renormalization flow, one can go down from $M_\calU$ to meet a new scale $\Lambda_\calU$ 
where the scale invariance emerges.
It appears through the dimensional transmutation.
Below $\Lambda_\calU$ the theory is matched onto the above interaction with
the new unparticle operator $\calO_\calU$ as
\begin{equation}
C_\calU\frac{\Lambda_\calU^{d_{\rm UV}-\dU}}{M_\calU^{d_{SM}+d_{\rm UV}-4}}
\calO_{SM}\calO_{\calU}~,
\end{equation}
where $\dU$ is the scaling dimension of $\calO_\calU$ and $C_\calU$ is the
matching coefficient.
\par
So far there have been a lot of investigations about unparticles in every respect \cite{Kingman,U}.
Among them is the spin-2 unparticle effects, or ungravity \cite{Goldberg,Das}.
Ungravity is induced by a traceless tensor unparticle operator $\calO_{\mu\nu}$ 
which couples to the energy-momentum tensor $T^{\mu\nu}$
\begin{equation}
\kappa_*\frac{1}{\Lambda_\calU}\sqrt{g}T^{\mu\nu}\calO_{\mu\nu}~,
\end{equation}
where $\kappa_*=\Lambda_\calU^{-1}(\Lambda_\calU/M_\calU)^{d_{\rm UV}}$ 
and $g$ is the determinant of the metric.
One of the most important result of ungravity is the power law correction to the
Newtonian gravitational potential, of type $\sim (1/r)^{2\dU-1}$, resulting in the enhancement of gravity.
Thus in the regime where the unparticle effect is strong, it is possible that black holes would be formed \cite{Mureika}.
This kind of ``$\calU$-enhanced`` black hole is very similar to the mini black holes in the extra-dimension scenarios 
\cite{ADD,XDBH,Kanti}. 
Properties of the extra-dimensional black holes have been widely studied.
One of them is the quasinormal modes of the $D$-dimensional black holes \cite{Konoplya,KK}.
\par
The quasinormal modes of black holes are  a phase of damping oscillations of external fields or the metric itself 
which perturbs the black holes \cite{Nollert, Kokkotas}.
This phase is governed by the complex quasinormal frequencies of the fields.
The real part of the quasinormal frequency drives the field oscillations while the imaginary part is responsible for the damping.
For a field whose time dependence is $\sim e^{-i\omega t}$ where $\omega$ is the frequency, the imaginary part of $\omega$
must be negative to guarantee damping with time.
(For rotating black holes the imaginary part could be positive in the case of superradiance where the rotation energy is extracted. 
See \cite{BHB}.)
The quasinormal modes of a black hole contain characteristic features of the black hole, 
so it is quite challenging to examine the quasinormal modes of the $\calU$-enhanced black holes and compare them with 
those of the extra-dimensional black holes.
\par
One way of distinguishing the extra-dimensional black holes and the $\calU$-enhanced black holes is to study the aspects
of Hawking radiations of both types of black holes \cite{Mureika}.
In this paper we propose that the quasinormal modes are very useful ''fingerprints`` to do the work.
Especially, the constraints by the unitarity on $\dU$ \cite{Grinstein} are so strong.
According to \cite{Grinstein}, unitarity requires $\dU\ge 4$ for tensor unparticles.
In this region it will be shown that some modes of field oscillations have ${\rm Im}~\omega>0$, 
which is unphysical because the field is amplified with time, not damped.
Consequently some of the quasinormal modes are forbidden for the unparticle black holes, 
quite contrary to those of extra-dimensional black holes.
\par
In the next Section, the $\calU$-enhanced black holes are described and the corresponding geometry is given.
With the induced metric, we construct the master equations which govern the propagation of fields under the ungravity backgrounds.
In Sec.\ III the results for the quasinormal modes of scalar, vector, and Dirac fields are provided.
To get the quasinormal frequencies we adopt the WKB approximation up to the 3rd order.
We conclude in Sec.\ IV.
\section{$\cal U$-enhanced black holes and master equations}
The ungravity effects by tensor unparticles result in the modification of the
gravitational potential as \cite{Mureika}
\begin{equation}
\Phi_\calU(r)
=\Phi_N(r)\left[1+\Gamma_\calU\left(\frac{R_*}{r}\right)^{2\dU-2}\right]~,
\label{PhiU}
\end{equation}
where
\begin{eqnarray}
\Gamma_\calU&=&
\frac{2}{\pi^{2\dU-1}}
\frac{\Gamma(\dU+\frac{1}{2})\Gamma(\dU-\frac{1}{2})}{\Gamma(2\dU)}~,\\
R_*&=&
\frac{1}{\Lambda_\calU}\left(\frac{\MPl}{\Lambda_\calU}\right)^{\frac{1}{\dU-1}}
\left(\frac{\Lambda_\calU}{M_\calU}\right)^{\frac{\dUV}{\dU-1}}~.
\end{eqnarray}
Here $\Phi_N(r)=-GM/r$ is the Newtonian potential.
We can consider the unparticle effects as a modification to the Schwarzschild metric
as follows \cite{Mureika}:
\begin{equation}
-g_{00}=1+2\Phi_\calU(r),~~~g_{11}=\frac{1}{1+2\Phi_\calU(r)}~.
\end{equation}
Thus the metric is of the form
\begin{equation}
ds^2=-h_\calU(r)dt^2+\frac{1}{h_\calU(r)}dr^2+r^2d\Omega^2~,
\label{metric}
\end{equation}
where
\begin{equation}
h_\calU(r)=
1-\frac{2GM}{r}\left[1+\Gamma_\calU\left(\frac{R_*}{r}\right)^{2\dU-2}\right]~.
\label{hU}
\end{equation}
For large $r$, $\Phi_\calU$ simply becomes the Newtonian potential. 
But for sufficiently small $r$, the unparticle effects dominate.
In this case the potential looks like that of a higher-dimensional gravity.
Since the higher-dimensional potential behaves as $\Phi\sim (1/r)^{D-3}$ in $D$-dimensions, 
one can easily find the correspondence
\begin{equation}
 2\dU-1=D-3~.
\label{dUn}
\end{equation}
In this analysis we will consider only the region where the ungravity effect is strong.
In this region, $h_\calU$ is approximately
\begin{equation}
 h_\calU(r)\approx 
1-\frac{2GM}{r}\Gamma_\calU\left(\frac{R_*}{r}\right)^{2\dU-2}~,
\label{hUapp}
\end{equation}
which resembles the Schwarzschild geometry in $D=2\dU+2$ dimensions.
\par
Now we consider the propagation of particles through the background of Eq.\ (\ref{metric}).
A field of spin $s=0, 1/2, 1$ is factorized as \cite{Kanti,KK}
\begin{equation}
 \Psi_s(t,r,\theta,\phi)=e^{-i\omega t}e^{im\phi}\Delta^{-s}R_s(r)S_s(\theta)~,
\end{equation}
where $\Delta=h(r)r^2$.
The radial and angular functions satisfy the ``master equations''
\begin{equation}
 \Delta^s\frac{d}{dr}\left(\Delta^{1-s}\frac{dR_s}{dr}\right)
+\left(\frac{\omega^2r^2}{h}+2is\omega r-\frac{is\omega r^2h'}{h}-{\tilde\lambda}\right)R_s=0~,
\label{radial}
\end{equation}
and
\begin{equation}
 \frac{1}{\sin\theta}\frac{d}{d\theta}\left(\sin\theta\frac{dS_s}{d\theta}\right)
+\left(-\frac{2ms\cot\theta}{\sin\theta}-\frac{m^2}{\sin^2\theta}+s-s^2\cot^2\theta+\lambda\right)S_s=0~,
\end{equation}
where
\begin{equation}
 {\tilde\lambda}=\lambda+2s=\ell(\ell+1)-s(s-1)~.
\end{equation}
The radial equation (\ref{radial}) can be written as a form of a one-dimensional Schr\"odinger equation \cite{Khanal, KK}
\begin{equation}
 \left(\frac{d^2}{dr_*^2}+\omega^2\right)Z_s=V_sZ_s~,
\end{equation}
where
\begin{equation}
 R_s=r^{2(s-1/2)}Y_s~, ~~~Y_s=hZ_s+2i\omega\left(\frac{d}{dr_*}-i\omega\right)Z_s~
\end{equation}
and $dr_*=dr/h$ is the so-called ``tortoise'' coordinate.
Here the effective potential $V_s$ takes different forms for its spin $s$ as follows:
\begin{eqnarray}
 V_{s=0}&=&h\left[\frac{\ell(\ell+1)}{r^2}+\frac{h'}{r}\right]~,\\
 V_{s=1}&=&\frac{h}{r^2}\ell(\ell+1)~,\\
 V_{s=1/2}&=&hk\left[\frac{k}{r^2}\mp\frac{d}{dr}\left(\frac{\sqrt{h}}{r}\right)\right]~,
\end{eqnarray}
where $k=\sqrt{\ell(\ell+1)+1/4}=1, 2, 3, \cdots$.

\section{Quasinormal modes}
The quasinormal modes satisfy the boundary conditions
\begin{equation}
 \Psi_s(r_*)\approx C_\pm e^{\pm i\omega r_*}~~~{\rm as} ~~~r_*\to\pm\infty~.
\end{equation}
In the original coordinate $r$, $r_*=-\infty$ is the event horizon and $r_*=+\infty$ is spatial infinity.
Thus the quasinormal mode is incoming at the event horizon and outgoing at infinity.
Note that the infinity is implied in the sense that the approximation of Eq.\ (\ref{hUapp}) is hold.
For example, the 2nd and the 3rd terms of Eq.\ (\ref{hU}) become comparable when $r\sim 10^{33} r_S\sim 10^3 r_\calU$ for $\dU=6$
(and gets larger for smaller $\dU$),
where $r_S=2GM$ is the Schwarzschild radius and 
\begin{equation}
 h_\calU(r_\calU)\approx 1-\frac{2GM}{r_\calU}\Gamma_\calU\left(\frac{R_*}{r_\calU}\right)^{2\dU-2}=0~.
\end{equation}
Current analysis will be done within this range of validity for simplicity.
\par
To get the quasinormal mode frequencies, one can use the well-known WKB approximation since the effective potential 
$V_s$ has the positive-definite potential barrier.
Up to the 3rd order of accuracy, the frequency is given by
\begin{equation}
 i\frac{\omega^2-V_0}{\sqrt{-2V_0''}}-L_2-L_3=n+\frac{1}{2}~,
\label{WKB3}
\end{equation}
where $V_0$ is the maximum height of $V$, and $V_0''$ is the second derivative of $V_s$ with respect to $r_*$ at the point $r_0$
which maximizes $V_s$, $V_s(r_0)=V_0$, and $n$ is the overtone number.
Here $L_{2, 3}$ are the WKB corrections and their explicit forms can be found in \cite{Iyer}.
One can add higher order terms to Eq.\ (\ref{WKB3}) as in \cite{KK,Konoplya}.
\par
Using Eq.\ (\ref{WKB3}), we provide the quasinormal mode frequencies in Tables I, II, and III.
In the analysis, we put $M_\calU=10$ TeV, $\Lambda_\calU=1$ TeV, $M=5$ TeV, and $\dUV=1$.
\begin{table}
\begin{tabular}{cc|cc}\hline\hline
$\dU=4$ &$\omega$ &$\dU=5$ &$\omega$ \\\hline\hline
$(\ell, n)=(0, 0)~~~$ &$0.116019+2.13604i$ &$(\ell, n)=(0, 0)~~~$ &$0.399872+2.86104i$ \\\hline
$(\ell, n)=(1, 0)~~~$ &$0.173311+1.81192i$ &$(\ell, n)=(1, 0)~~~$ &$0.530601+2.6313i$ \\
$(\ell, n)=(1, 1)~~~$ &$0.541964+5.17633i$ &$(\ell, n)=(1, 1)~~~$ &$0.852281+6.8386i$ \\\hline
$(\ell, n)=(2, 0)~~~$ &$0.609546-0.691852i$ &$(\ell, n)=(2, 0)~~~$ &$0.678472+1.81891i$ \\
$(\ell, n)=(2, 1)~~~$ &$1.1295+4.59508i$ &$(\ell, n)=(2, 1)~~~$ &$2.41065+6.66656i$ \\
$(\ell, n)=(2, 2)~~~$ &$3.5565+8.48243i$ &$(\ell, n)=(2, 2)~~~$ &$5.69487+11.9428i$ \\\hline
$(\ell, n)=(3, 0)~~~$ &$2.14206-0.874824i$ &$(\ell, n)=(3, 0)~~~$ &$1.79741-0.820522i$ \\
$(\ell, n)=(3, 1)~~~$ &$0.244724-3.47487i$ &$(\ell, n)=(3, 1)~~~$ &$1.62507+5.25536i$ \\
$(\ell, n)=(3, 2)~~~$ &$2.0898+6.93238i$ &$(\ell, n)=(3, 2)~~~$ &$5.47658+10.4901i$ \\
$(\ell, n)=(3, 3)~~~$ &$4.98341+10.7721i$ &$(\ell, n)=(3, 3)~~~$ &$10.4015+16.4574i$ \\\hline\hline
\end{tabular}
\caption{WKB3 approximation for the quasinormal frequencies for scalar fields in units of $1/r_\calU$.}
\label{scalar}
\end{table}
\begin{table}
\begin{tabular}{cc|cc}\hline\hline
$\dU=4$ &$\omega$ &$\dU=5$ &$\omega$ \\\hline\hline
$(\ell, n)=(1, 0)~~~$ &$0.19116+1.24064i$ &$(\ell, n)=(1, 0)~~~$ &$0.523096+1.6981i$ \\
$(\ell, n)=(1, 1)~~~$ &$2.06499+4.31569i$ &$(\ell, n)=(1, 1)~~~$ &$3.5286+5.92237i$ \\\hline
$(\ell, n)=(2, 0)~~~$ &$1.11899-0.854651i$ &$(\ell, n)=(2, 0)~~~$ &$0.83325-1.03127i$ \\
$(\ell, n)=(2, 1)~~~$ &$0.525274+3.55141i$ &$(\ell, n)=(2, 1)~~~$ &$1.65328+4.80792i$ \\
$(\ell, n)=(2, 2)~~~$ &$2.7902+6.77041i$ &$(\ell, n)=(2, 2)~~~$ &$5.17625+9.3328i$ \\\hline
$(\ell, n)=(3, 0)~~~$ &$2.0822-0.881647i$ &$(\ell, n)=(3, 0)~~~$ &$2.01102-1.02585i$ \\
$(\ell, n)=(3, 1)~~~$ &$0.690689-3.12684i$ &$(\ell, n)=(3, 1)~~~$ &$0.303258+4.13687i$ \\
$(\ell, n)=(3, 2)~~~$ &$1.24996+6.04807i$ &$(\ell, n)=(3, 2)~~~$ &$3.26883+8.23007i$ \\
$(\ell, n)=(3, 3)~~~$ &$3.64909+9.32242i$ &$(\ell, n)=(3, 3)~~~$ &$6.97334+12.8731i$ \\\hline\hline
\end{tabular}
\caption{WKB3 approximation for the quasinormal frequencies for vector fields in units of $1/r_\calU$.}
\label{vector}
\end{table}
\begin{table}
\begin{tabular}{cc|cc}\hline\hline
$\dU=4$ &$\omega$ &$\dU=5$ &$\omega$ \\\hline\hline
$(\ell, n)=(1, 0)~~~$ &$0.274179+1.84419i$ &$(\ell, n)=(1, 0)~~~$ &$0.62748+2.51579i$ \\
$(\ell, n)=(1, 1)~~~$ &$0.994088+4.83761i$ &$(\ell, n)=(1, 1)~~~$ &$1.71823+6.34767i$ \\\hline
$(\ell, n)=(2, 0)~~~$ &$0.467238-1.30984i$ &$(\ell, n)=(2, 0)~~~$ &$0.126116-1.95711i$ \\
$(\ell, n)=(2, 1)~~~$ &$0.104486+4.52932i$ &$(\ell, n)=(2, 1)~~~$ &$0.56586+5.96395i$ \\
$(\ell, n)=(2, 2)~~~$ &$1.09812+7.72804i$ &$(\ell, n)=(2, 2)~~~$ &$1.9171+10.0892i$ \\\hline
$(\ell, n)=(3, 0)~~~$ &$1.45664-0.896816i$ &$(\ell, n)=(3, 0)~~~$ &$0.949156-1.26542i$ \\
$(\ell, n)=(3, 1)~~~$ &$0.363197-4.20139i$ &$(\ell, n)=(3, 1)~~~$ &$0.0248592-5.698i$ \\
$(\ell, n)=(3, 2)~~~$ &$0.727909+0.746487i$ &$(\ell, n)=(3, 2)~~~$ &$1.14576+9.86467i$ \\
$(\ell, n)=(3, 3)~~~$ &$2.25403+10.8068i$ &$(\ell, n)=(3, 3)~~~$ &$2.83694+14.119i$ \\\hline\hline
\end{tabular}
\caption{WKB3 approximation for the quasinormal frequencies for Dirac fields in units of $1/r_\calU$.}
\label{Dirac}
\end{table}
For scalar and vector fields, one can easily find the analytic form of $r_0$ which makes $V_s$ maximum.
\begin{eqnarray}
 r_0(s=0)&=&\left[\frac{A+\sqrt{A^2+32\ell(\ell+1)\dU(2\dU-1)}}{4\ell(\ell+1)}\right]^{\frac{1}{2\dU-1}}r_\calU~,
\label{r0s0}\\
 r_0(s=1)&=&\left(\frac{2\dU+1}{2}\right)^{\frac{1}{2\dU-1}}r_\calU~,
\label{r0s1}
\end{eqnarray}
where $A=(1+2\dU)[\ell(\ell+1)-2\dU+1]$.
Note that Eq.\ (\ref{r0s1}) is exactly same as $r_0(s=1)$ for the brane-localized vector fields \cite{KK}, 
considering the correspondence Eq.\ (\ref{dUn}).
\par
As for Dirac fields ($s=1/2$) it is quite difficult to extract the analytic form of $r_0$, so we find $r_0$ by the numerical methods.
\par
In Figures 1 and 2, the imaginary part of $\omega$ is plotted as a function of $\dU$ by using Eqs.\  (\ref{r0s0}) and (\ref{r0s1}).
\begin{figure}
\begin{tabular}{lr}
\includegraphics{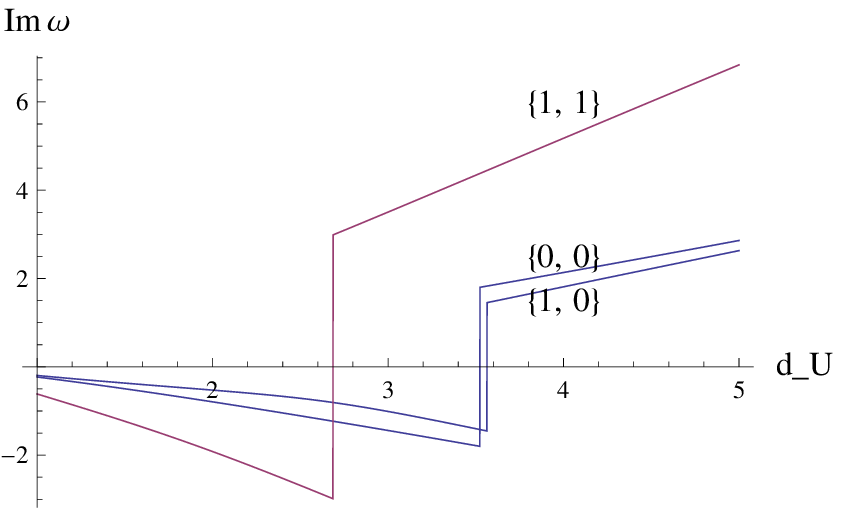}&\includegraphics{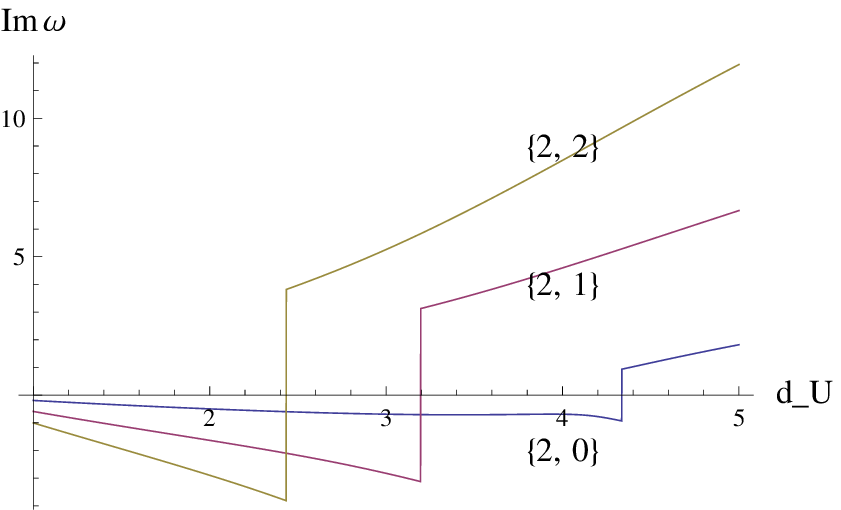}\\
\includegraphics{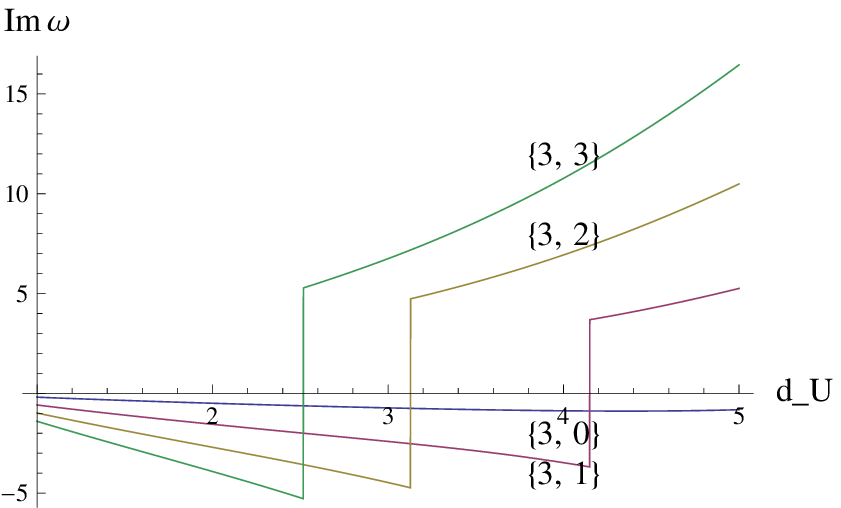}&
\end{tabular}
\caption{\label{S}Imaginary parts of $\omega$ for scalar fields as a function of $\dU$.
Numbers are shown to denote the modes $\{\ell, n\}$.}
\end{figure}
\begin{figure}
\begin{tabular}{lr}
\includegraphics{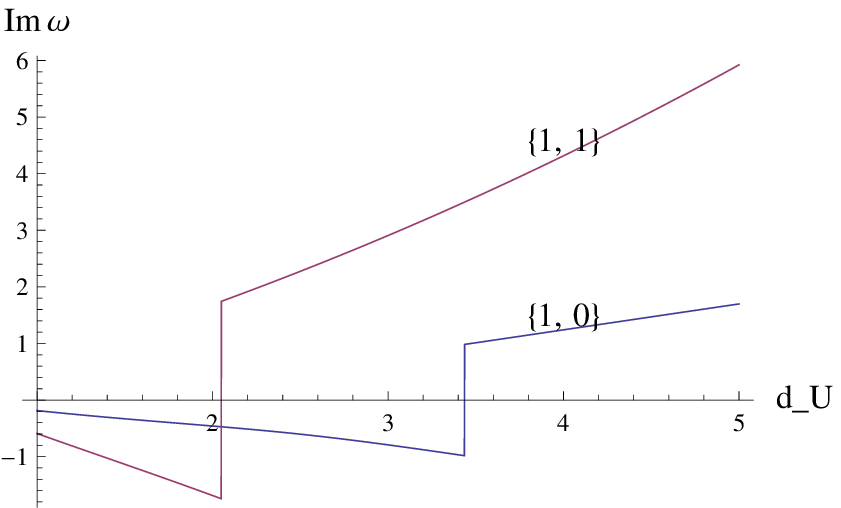}&\includegraphics{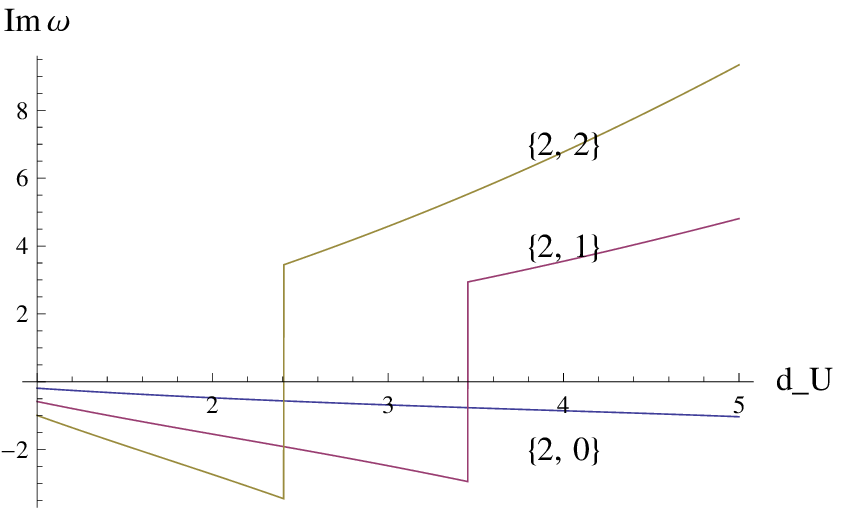}\\
\includegraphics{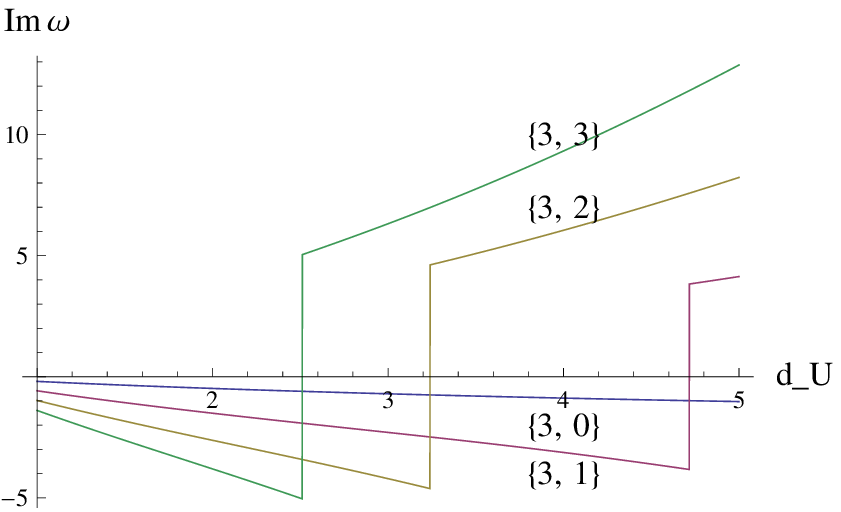}&
\end{tabular}
\caption{\label{V}Imaginary parts of $\omega$ for vector fields.}
\end{figure}
In Fig.\ \ref{D}, ${\rm Im}~\omega$ is plotted numerically in the range of $4\le\dU\le 5$.
Note that the unitarity-allowed region is $\dU\ge 4$, and ${\rm Im}~\omega$ is not always negative in this range.
Thus Figures 1, 2, and 3 (also Tables I, 2, and III) show that not all the modes of $(\ell, n)$ exhibit the quasinormal modes for 
ungravity background.
\begin{figure}
\begin{tabular}{lr}
\includegraphics{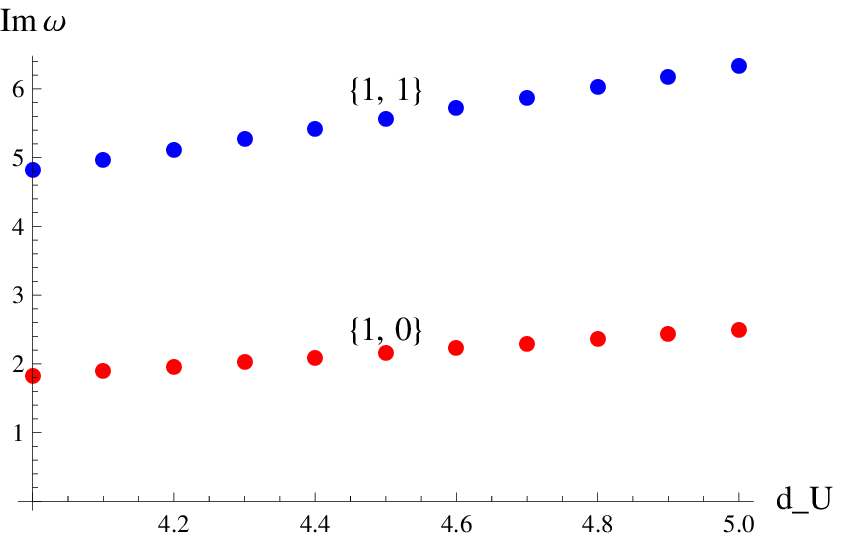}&\includegraphics{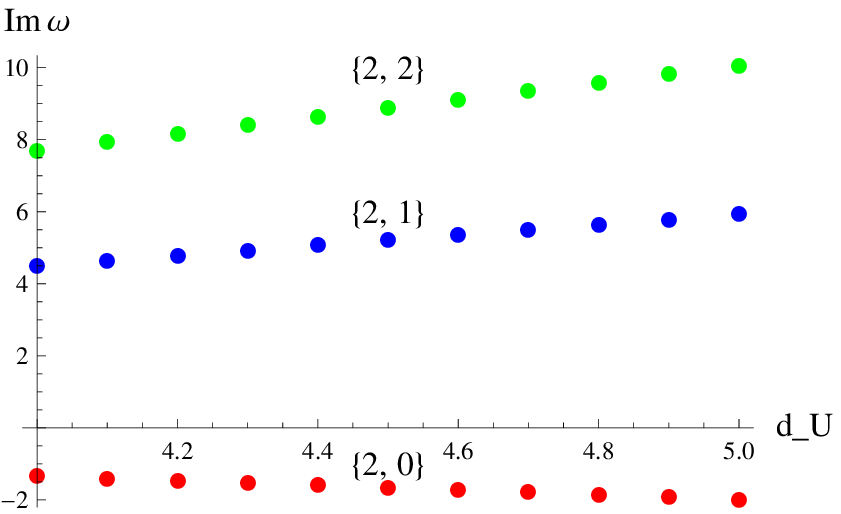}\\
\includegraphics{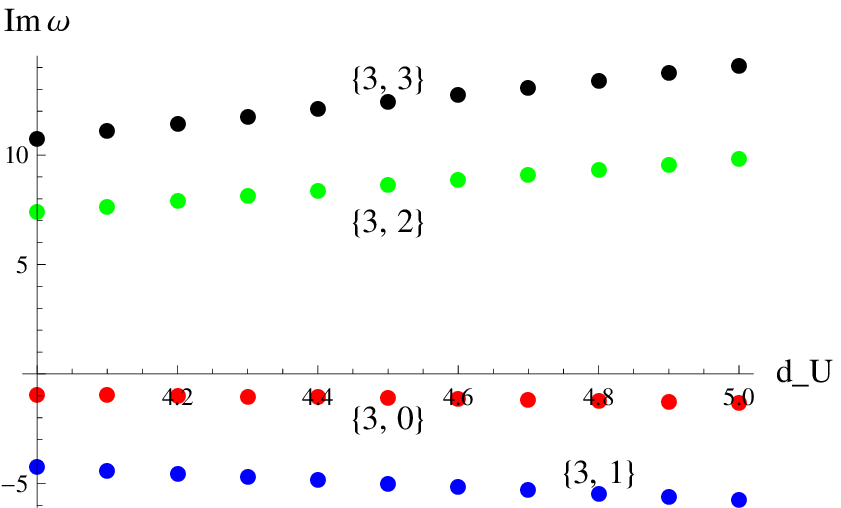}&
\end{tabular}
\caption{\label{D}Imaginary parts of $\omega$ for Dirac fields.
Numbers are shown to denote the modes $\{k, n\}$.}
\end{figure}
For example, scalar modes of $(\ell, n)=(0, 0), (1, 0), (1, 1), (2, 1), (2, 2), (3, 2), (3, 3)$ have no quasinormal modes for
$\dU\ge 4$ (see Fig.\ \ref{S}).
This is quite a distinctive point compared with the brane-localized model \cite{KK}.
For scalar $(2, 0)$ mode, the quasinormal mode is allowed only for $4\le\dU\le 4.336$, 
and for scalar $(3, 1)$, the allowed region is $4\le\dU\le 4.149$.
The scalar $(3, 0)$ mode is a quasinormal mode for all the range of $\dU$ considered in this analysis.
\par
According to Fig.\ \ref{V}, vector fields show much the same behavior as scalar ones except that 
quasinormal $(2, 0)$ is allowed for $4\le\dU$ and quasinormal $(3, 1)$ is for $4\le\dU\le 4.716$.
As for Dirac fields, quasinormal $(k, n)=(2, 0), (3, 0), (3, 1)$ modes are allowed for $4\le\dU\le 5$ and other modes are not.
\par
\begin{figure}
\includegraphics{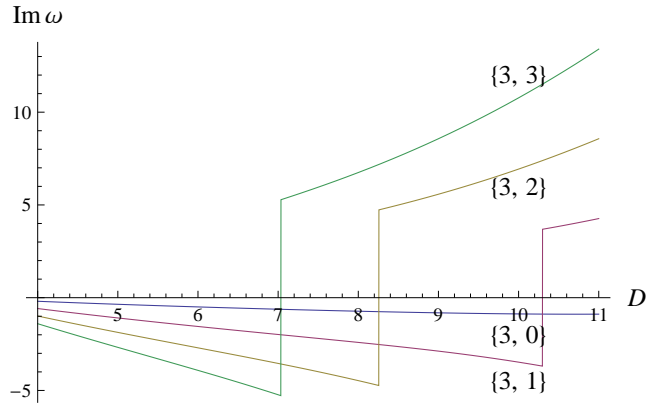}
\caption{\label{cf}Imaginary parts of $\omega$ for scalar fields near higher-dimensional black hole backgrounds.}
\end{figure}

The imaginary parts of the quasinormal frequencies of brane-localized fields near higher-dimensional black holes show 
similar behaviors to Figs. \ref{S}-\ref{D}.
Actually, when $\dUV=1$, Eq.\ (\ref{hUapp}) becomes
\begin{equation}
 h_\calU(r)\approx 1-\frac{2M\Gamma_\calU}{M_\calU^2\Lambda_\calU^{2\dU-2}}
\left(\frac{1}{r}\right)^{2\dU-1}~,
\end{equation}
which looks much like 
\begin{equation}
 h(r)=1-\frac{\mu}{r^{D-3}}~,
\label{hD}
\end{equation}
where $\mu$ is some mass parameter, in $D$-dimensional black hole backgrounds \cite{KK}, considering Eq.\ (\ref{dUn}).
For example, the imaginary parts of $\omega$ of scalar fields for some modes with Eq.\ (\ref{hD}) 
are shown in Fig.\ \ref{cf}.
In Fig.\ \ref{cf}, all the modes of $\ell=3$ are quasinormal when $D\lesssim 7$, 
which corresponds to $\dU\lesssim 2.5$ (see Fig.\ \ref{V}) for ungravity backgrounds.
\section{Conclusions}
In this work, we have analyzed the quasinormal modes of particles in the unparticle-enhanced black hole background.
The effects of tensor unparticles look like a slight modification to the Schwarzschild metric of a spacetime with
$(2\dU+2)$ dimensions, and the black hole formation might be possible at the LHC without any extra dimensions.
We have shown that the quasinormal spectrum is a good probe for the properties of the black hole background.
Basically unparticle-enhanced black holes can be considered as $(2\dU+2)$-dimensional black holes.
But there are two big differences.
One is that $\dU$ can be any real number.
For example if one finds a fractional number of extra dimensions by examining the quasinormal modes, 
then the background geometry is given by unparticles.
The other is that many quasinormal modes are forbidden in the unparticle background by the unitarity bound.
This is a crucial feature to distinguish unparticles from extra dimensions.
\par
Current work is done for massless particles, but the extension to the massive fields is also promising as was done
for the conventional black hole backgrounds.
And including higher WKB corrections is quite straightforward to improve the accuracy.
\begin{acknowledgments}
The author thanks P. Kanti and R.A. Konoplya for invaluable discussions and helps.
This work was supported by the Basic Science Research Program through the National Research Foundation of Korea (NRF) 
funded by the Korean Ministry of Education, Science and Technology (2009-0088396).
\end{acknowledgments}

\end{document}